\documentclass[aps,10pt,twocolumn,longbibliography,superscriptaddress]{revtex4-1}
\usepackage{graphicx,color,multirow,bm,braket,amsmath,soul,epstopdf}

\newcommand{\rr}{\mathbf{r}}

\def\bk{{\bf k}}
\def\bq{{\bf q}}
\def\ve{\varepsilon}
\def\e{\epsilon}
\DeclareMathAlphabet{\mathcal}{OMS}{cmsy}{m}{n}

\begin{document}
 
\title{Many-body calculations of plasmon and phonon satellites in angle-resolved photoelectron
spectra using the cumulant expansion approach}
\author{Fabio Caruso}
\affiliation{Institut f\"ur Physik and IRIS Adlershof, Humboldt-Universit\"at zu Berlin, Berlin, Germany}
\author{Carla Verdi}
\affiliation{Department of Materials, University of Oxford, Parks Road, Oxford OX1 3PH, United Kingdom}
\author{Feliciano Giustino}
\email{feliciano.giustino@materials.ox.ac.uk}
\affiliation{Department of Materials, University of Oxford, Parks Road, Oxford OX1 3PH, United Kingdom}
\affiliation{Department of Materials Science and Engineering, Cornell University, Ithaca, New York, 14850, USA}

\begin{abstract}
The interaction of electrons with crystal lattice vibrations (phonons) and collective charge-density 
fluctuations (plasmons) influences profoundly the spectral properties of solids revealed by photoemission 
spectroscopy experiments. Photoemission satellites, for instance, are a prototypical example of quantum 
emergent behavior that may result from the strong coupling of electronic states to plasmons and phonons. 
The existence of these spectral features has been verified over energy scales spanning several orders of 
magnitude (from 50~meV to 15-20~eV) and for a broad class of compounds such as simple metals, semiconductors, 
and highly-doped oxides. During the past few years the cumulant expansion approach, alongside with the $GW$ 
approximation and the theory of electron-phonon and electron-plasmon coupling in solids, has evolved into 
a predictive and quantitatively accurate approach for the description of the spectral signatures of 
electron-boson coupling entirely from first principles, and it has thus become the state-of-the-art 
theoretical tool for the description of these phenomena. In this chapter we introduce the fundamental 
concepts needed to interpret plasmon and phonon satellites in photoelectron spectra, and we review 
recent progress on first-principles calculations of these features using the cumulant expansion method.
\end{abstract}

\maketitle

\section{Introduction}

The emergence of satellites in photoemission spectroscopy provides direct evidence of the electronic 
coupling to bosonic excitations in solids. Satellites are spectral features that reflect the simultaneous 
excitation of a hole and of a boson, and they are separated from the quasiparticle peak by a multiple of 
the boson energy. The origin of these features may be understood based on simple considerations on the 
energy scales involved in the photoemission process. When a photon with energy $\hbar\omega$ is absorbed 
by an electron with binding energy $\e_i$, if no boson modes are excited in the system, energy conservation 
requires the condition $\hbar \omega = \e_i + \Phi + E_\textup{kin}$ to be satisfied, where $\Phi$ is the 
work-function of the system and $E_\textup{kin}$ is the kinetic energy of the photo-emitted electron. In 
photoemission, by measuring $E_\textup{kin}$ and $\Phi$, the electron binding energy can thus be inferred. 
If, in addition to the creation of a hole, a fraction of the absorbed photon energy is transferred to the 
system in the form of bosonic modes, such as plasmons and phonons, the energy conservation condition is 
modified as follows: $ E_\textup{kin}= \hbar \omega - \e_i - {n}E_\textup{b} - \Phi$, where $E_\textup{b}$ 
is the energy of the boson and $n$ an integer. Since $\hbar \omega$ and $\Phi$ are constants, the kinetic 
energy distribution of the photo-emitted electrons will be peaked at the energies corresponding to (i) the 
binding energy of electrons $\e_i$ and (ii) the sum of the binding and boson energies $\e_i + {n}E_\textup{b}$, 
and it may thus provide direct information regarding the coupling of electrons to bosonic modes in solids.

The presence of satellites in the photoemission spectra of solids was first predicted by a theoretical analysis 
of the spectral function of the homogeneous electron gas by \citep{Lundqvist1967,HEDIN19701,Langreth1970}, 
and subsequently verified experimentally for the core electrons of simple metals \citep{Baer1973}. Recently, 
the availability of energy resolutions of the order of 25-50~meV in angle-resolved photoelectron 
spectroscopy (ARPES) made it possible to observe new low-energy signatures of electron-boson coupling 
in experiments. In particular, high-resolution ARPES measurements of graphene by \citep{Bostwick2010} 
have revealed plasmon-induced satellite structures with characteristic energies of the order of $\sim1$~eV. 
More recently, polaronic satellites at energies of the order $\sim100$~meV from the band edges have been 
observed in doped oxides, for example by \citep{Moser2013}, \citep{Baumberger2016}. At variance with 
valence-plasmon satellites, which typically appear at energies between $5$ and $15$~eV below the Fermi 
energy and have been known since the early days of photoemission spectroscopy, low-energy satellites are 
a manifestation of the coupling between low-energy bosonic modes and electronic carriers near the band edges.
In addition to the formation of satellites, the coupling to {bosons} may lead to the emergence of 
photoemission kinks \citep{Lanzara2001,Damascelli2003} and to a renormalization of energy levels 
\citep{Logothetidis1992,GiustinoPRL2010,Ponce2015} and carrier lifetimes \citep{Eiguren2002,Park2007}.

In this chapter we will discuss the state-of-the-art techniques for the description of plasmon and polaron 
satellites and their application to the prediction and interpretation of photoemission spectroscopy experiments.

\section{The localized-electron model}\label{sec:model}

To illustrate how the interaction between electrons and bosons may lead to the emergence 
of satellites in photoemission spectra, we consider in the following the exactly solvable 
model of a ``localized electron'' in a solid interacting with a boson bath. The latter can 
be regarded as a set of phonons, plasmons, or any other bosonic excitations that may be 
approximately represented as a set of uncoupled harmonic oscillators. The localized electron 
is assumed to be dispersionless, that is, its energy $\ve$ is independent of the crystal 
momentum, and its interaction with other electrons in the system is neglected. 
Instances in which the electron energy levels exhibit a weak dependence on momentum, and 
can thus be approximated as non-dispersive, are for example core electrons in solids, localized 
impurity levels, and $4f$ electrons. On the other hand, electron-electron interactions 
are typically strong and non-negligible in three-dimensional solids, which poses limitations to 
the applicability of this model to real physical systems. 
This simplified model, however, is remarkably successful in describing the emergence of 
bosonic satellites in the spectral properties, and is in good qualitative agreement with more 
advanced theories whereby the electron-electron interaction is accounted for. This is demonstrated, 
for instance, by the generalization of the localized electron model reported by \citep{Langreth1970}.

The localized electron model is described by the following electron-boson coupling Hamiltonian: 
\begin{align}
\hat{H} &= \hat{H}_{\rm e} + \hat{H}_{\rm b} + \hat{H}_{\rm int} \nonumber \\
 &= \ve\, \hat{c}^\dagger \hat{c}  + \sum_\bq \hbar\omega_\bq \hat{b}_\bq^\dagger \hat{b}_\bq 
 + \sum_\bq g_\bq \hat{c}^\dagger \hat{c}\,(\hat{b}_\bq + \hat{b}_{-\bq}^\dagger), \label{eq:Heb}
\end{align}
where $\hat{c}^\dagger$ and $\hat{c}$ are fermionic creation and 
annihilation operators for the localized electron, respectively, which satisfy 
the ordinary anti-commutation relations. 
Similarly, the operators $\hat{b}_{\bf q}^\dagger$ and $\hat{b}_{\bf q}$ respectively create  and annihilate
a boson with energy $\hbar \omega_{\bf q}$ and momentum ${\bf q}$ and satisfy bosonic commutation relations. 
The absence of two-particle interaction terms in the Hamiltonian reflects the fact 
that both electron-electron and boson-boson interactions are neglected.
The localized electron interacts with the boson bath with the coupling strength $g_{\bf q}$.

As we are primarily interested in the effects of the electron-boson interaction on 
the photoemission intensity, the relevant quantity that we want to compute is the 
electron spectral function: 
\begin{equation}\label{eq:A}
A(\omega) = {-}\frac{1}{\pi} {\rm Im}\,G^{\rm ret}(\omega),
\end{equation}
with the single-particle retarded Green's function $G^{\rm ret}$ defined as:
\begin{equation}
G^{\rm ret}(t) = -i \left\langle \Psi_0 \right|\lbrace\hat{c}(t), \hat{c}^\dagger(0)\rbrace 
\left| \Psi_0 \right\rangle \theta(t) ,
\end{equation}
where $\lbrace,\rbrace$ denotes the anticommutator, 
$\Psi_0$ the electronic ground state, and the time-dependence of the 
operators is accounted for in the Heisenberg picture. 
As shown by \citep{Langreth1970}, the Green's function associated to the Hamiltonian in Eq.~\eqref{eq:Heb}
can be calculated exactly. In fact, by applying a unitary transformation, Eq.~\eqref{eq:Heb} 
is recast in the form of a shifted harmonic oscillator Hamiltonian for which 
eigenvalues and eigenvectors are known \citep{Mahan2000}. 
The spectral function for a localized electron can thus be expressed as \citep{Langreth1970}: 
\begin{equation}\label{eq:Cmod}
A(\omega) = \sum_{n=0}^\infty \frac{e^{-a}a^n}{n!}
\delta(\hbar\omega - \ve - a \hbar \omega_{\rm b} + n \hbar \omega_{\rm b} ),
\end{equation}
where $a= \sum_{\bf q} g_{\bf q}^2 /(\hbar\omega_{\rm b})^2 $ and for simplicity 
the energy of the boson mode has been replaced by its average value $\hbar\omega_{\rm b}$.

In the small coupling limit, that is for ${g_{\bf q} \rightarrow 0}$, the spectral function reduces 
to the case of {a} non-interacting electron $A(\omega) = \delta(\hbar\omega-\ve)$, and the 
Dirac delta function is peaked at the quasiparticle energy. For finite coupling strengths, the 
structure of the spectral function in Eq.~(\ref{eq:Cmod}) reveals that the effect of the 
interaction between electrons and bosons on the spectral properties of the system is twofold. 
First, the quasiparticle energy of the localized electron is shifted by $a \hbar \omega_{\rm b}$. 
This process is analogous, for example, to the well-known band-gap renormalization of semiconductors 
and insulators due to the electron-phonon interaction {\citep{Allen1976,Giustino2017}} and it 
results from the {\it dressing} of the bare quasiparticle via the interaction with the boson modes. 
Second, the spectral function exhibits a series of additional features at lower energies which are 
separated from the quasiparticle {peak} by multiples of the boson energy $\hbar \omega_{\rm b}$. 
These spectral features arise from the simultaneous excitation of the localized electron and of 
one or more bosons with energy $\hbar \omega_{\rm b}$.

\begin{figure}[t]
\includegraphics[width=0.48\textwidth]{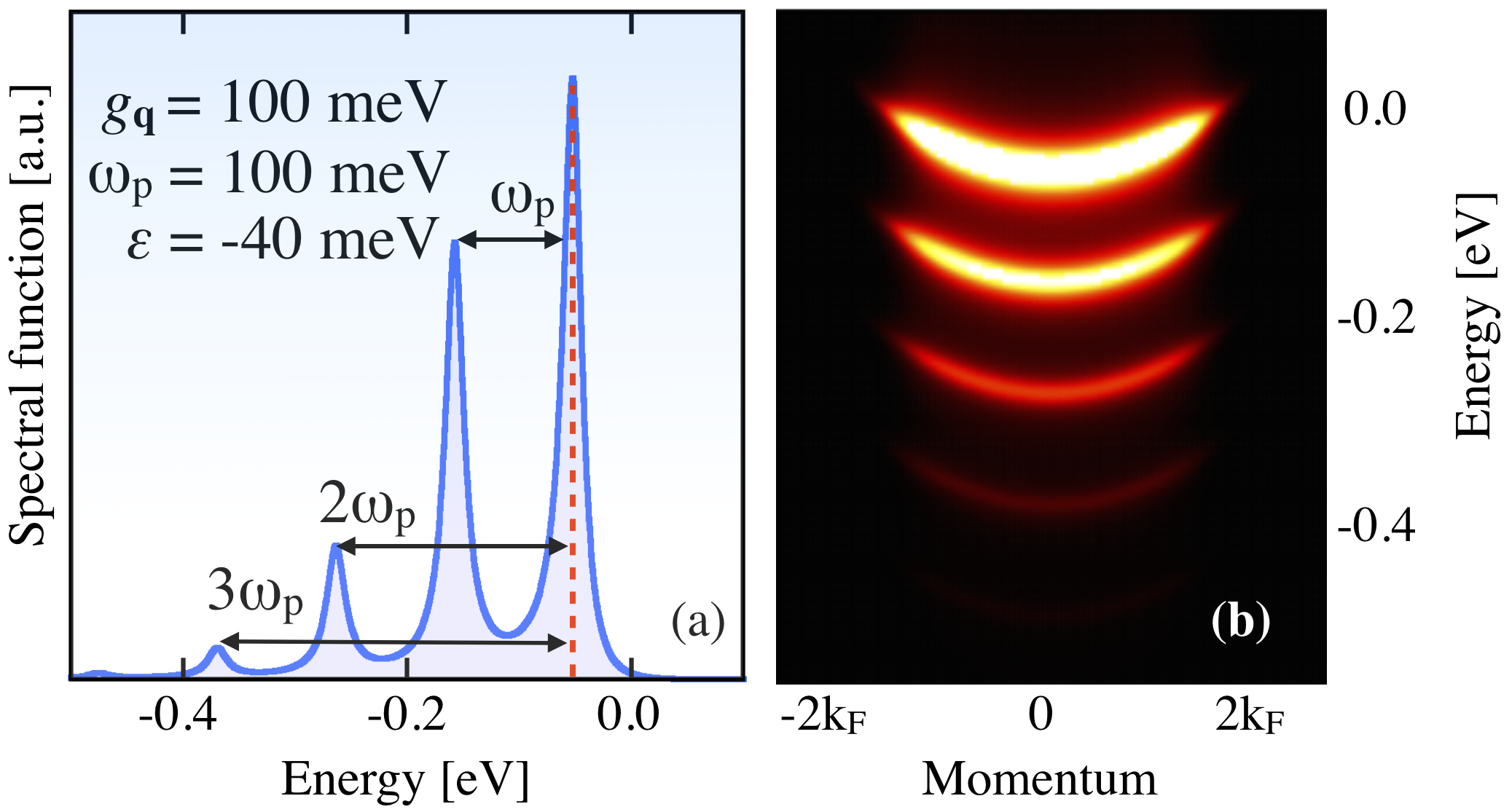}
\caption{\label{fig:Amod}(a) Spectral function of the localized electron model evaluated using Eq.~(\ref{eq:Cmod}) 
for a non-dispersive electron with binding energy $\ve$ coupled to a boson with frequency $\omega_{\rm b}$ 
with a coupling strength $g_{\bf q}$. (b) Spectral intensity map for an electron with parabolic band dispersion 
$\ve_{\bf k} = \ve + \hbar^2k^2/2m $. 
The Dirac $\delta$ functions in Eq.~\eqref{eq:Cmod} have been replaced by Lorentzian functions with a 20~meV 
broadening.}
\end{figure}

In Fig.~\ref{fig:Amod}(a) we show the spectral function obtained from Eq.~\eqref{eq:Cmod} considering 
$\ve = -40$~meV, $\hbar\omega_{\rm b} = 100$~meV, $ g_\bq = 100$~meV. A picture in closer agreement 
with angle-resolved photoemission spectroscopy is obtained when considering the case of dispersive electronic 
states{: Fig.~\ref{fig:Amod}(b)} illustrates the spectral intensity map for a parabolic band, obtained 
by replacing the electronic energy with $\ve_\bk = \ve + \hbar^2k^2/2m$ in Eq.~\eqref{eq:Cmod}. 
This simple generalization of the localized electron model illustrates that, in presence of non-trivial 
energy-wavevector dispersion relations, the energy of the satellite features induced by electron-boson 
coupling also acquires a dependence on the crystal momentum that follows closely the dispersion 
of the ordinary quasiparticle states. 
This phenomenon {translates into} the formation of plasmonic polaron bands due to electron-plasmon 
coupling \citep{Caruso/2015/PRL,Caruso/2015/PRB,Lischner/2015,Gumhalter/2016/PRB,Caruso2018} 
and polaron satellites due to electron-phonon coupling \citep{Moser2013,Baumberger2016,Verdi2017}
in the ARPES spectra of semiconductors and $n$-doped oxides, respectively.

Overall, the solution of the localized-electron model reveals that the spectral function of a system of 
interacting electrons and bosons, whereby the interaction is described by the last term of Eq.~\eqref{eq:Heb}, 
may exhibit a series of satellite structures, with binding energy blueshifted with respect to the main 
quasiparticle peak by multiples of the boson energy. Despite the simplicity of the model, this result 
provides a first indication that the coupling to plasmons and phonons in real systems, in which the coupling 
Hamiltonian assumes a similar form, may also induce the formation of satellite features for sufficiently 
strong coupling. 

\section{First-principles description of satellites in photoemission} \label{sec:theo}

Despite the different nature of plasmon and phonon {collective excitations} in solids, the many-body 
theory of electron-boson interaction represents the common playground to describe their coupling to 
electronic states, and to investigate the spectral fingerprints resulting from this interaction. The 
Hedin-Baym equations \citep{Giustino2017} based on many-body perturbation theory (MBPT) provide a formally 
exact framework to {investigate} the coupling to plasmons and phonons, and are the starting point for 
the theoretical description of satellites in photoemission spectra. The electron self-energy for the 
coupled electron-phonon system in the Migdal approximation, that is neglecting vertex corrections, is 
given by \citep{HEDIN19701,Giustino2017}:
\begin{equation}\label{eq:sigmatot}
\Sigma(\bk,\omega) = i\!\int\!\!\frac{d\bq}{\tilde\Omega}\frac{d\omega}{2\pi} \,G(\bk+\bq,\omega+\omega') 
[W_{\rm e}(\bq,\omega')+ W_{\rm ph}(\bq,\omega') ]
\end{equation}
where $\tilde\Omega$ is the reciprocal-space volume, $G$ is the single-particle Green's function and 
$W_{\rm e}$ ($W_{\rm ph}$) is the screened Coulomb interaction due to the electron-electron 
(electron-phonon) interaction. It can be shown that the electron-phonon part {may} be expressed as 
$W_{\rm ph}= W_{\rm e} D W_{\rm e}$ (in symbolic notation), where $D$ is the density-density 
correlation function for the nuclear fluctuations. Eq.~\eqref{eq:sigmatot} neglects the so-called 
Debye-Waller contribution to the self-energy, however this contribution is frequency-independent 
and therefore does not give rise to additional structure in the electron spectral 
function \citep{Giustino2017}.

If the nuclei are treated in the clamped-ion approximation, that is $W_{\rm ph}(\bq,\omega) = 0$, the 
ordinary $GW$ approximation is recovered. By expanding the Bloch wavefunctions in a basis set of plane waves, 
$\psi_{n\bk}(\rr)= \sum_{\bf G} c_{n\bk}({\bf G}) e^{i(\bk+{\bf G})\cdot\rr}$, the $GW$ self-energy 
{$\Sigma^{GW}$} can be expressed in a form more suitable for first-principles calculations of 
crystalline solids \citep{AULBUR2000}:
  \begin{multline}\label{eq:sigmapw}
  \Sigma_{n\bk}^{{GW}} (\omega) = \frac{i \hbar}{2\pi} \sum_{m{\bf G}{\bf G'}} 
  \int\!\frac{d\bq}{\Omega_{\rm BZ}}  M^{mn}_{\bf G}(\bk,\bq)^* M^{mn}_{\bf G'}(\bk,\bq)
  \\ \times \int \!d\omega' 
  \frac{v_{\bf G}(\bq) \epsilon^{-1}_{{\bf G},{\bf G'}} (\bq,\omega)}
  {\hbar\omega+\hbar\omega' { -\tilde{\ve}_{m\bk+\bq} }},
  \end{multline}
where $M^{mn}_{\bf G}(\bk,\bq)= \langle \psi_{m\bk+\bq} |e^{i(\bq+{\bf G})\cdot{\rr}}|\psi_{n{\bk}} \rangle$ 
are the optical matrix elements, $\Omega_{\rm BZ}$ is the volume of the Brillouin zone, and 
$v_{\bf G}(\bq)=e^2/\varepsilon_0|{\bf q + G}|^2 $ ($\varepsilon_0$ is the vacuum permittivity). We defined 
{$\tilde{\ve}_{m\bk+\bq}=\ve_{m\bk+\bq}-i\eta\,\mbox{sgn}(\ve_{m\bk+\bq})$, with $\eta$ 
a positive infinitesimal and $\ve_{m\bk+\bq}$ the Bloch electron energy relative to the chemical 
potential $\mu$.} The dielectric matrix $\epsilon_{{\bf G},{\bf G'}} (\bq,\omega)$ is related to the 
screened Coulomb interaction via $W_{{\bf G},{\bf G'}} (\bq,\omega)= v_{\bf G}(\bq) 
\epsilon^{-1}_{{\bf G},{\bf G'}} (\bq,\omega)$. 
In $GW$ calculations, the dielectric function is typically expressed as $\epsilon_{{\bf G},{\bf G'}} (\bq,\omega)
=\delta_{{\bf G},{\bf G'}} - v_{\bf G}(\bq) \chi^0_{{\bf G},{\bf G'}} (\bq,\omega)$, where 
$\chi^0_{{\bf G},{\bf G'}}$ is the independent-particle polarizability (see, e.g., \citep{AULBUR2000}). 

The second term in Eq.~\eqref{eq:sigmatot} represents the electron-phonon self-energy {$\Sigma^\textup{ep}$} 
in the Migdal approximation. Its expression in the basis of single-particle Bloch wavefunctions reads:
\begin{multline}\label{eq:sigma-eph}
\Sigma_{n\bk}^\textup{{ep}}(\omega)=\sum_{m\nu} \int\!\frac{d\bq}{\Omega_{\rm BZ}}|g_{mn\nu}(\bk,\bq)|^2 \,
\times \\  \left[ \frac{n_{\bq\nu}+ f_{m\bk+\bq}}{\hbar\omega -\ve_{m\bk+\bq}+\hbar\omega_{\bq\nu}+i\eta}+ 
 \frac{n_{\bq\nu}+1-f_{m\bk+\bq}} {\hbar\omega-\ve_{m\bk+\bq}-\hbar\omega_{\bq\nu}+i\eta} \right] ,
\end{multline}
where $n_{\bq\nu}$ and $f_{m\bk+\bq}$ are the Bose-Einstein and Fermi-Dirac distributions, respectively. 
Eq.~\eqref{eq:sigma-eph} {is derived after transforming} the frequency integration in 
Eq.~\eqref{eq:sigmatot} into a Matsubara summation to extend the formalism to finite temperatures, and 
{performing the integration} analytically {by using the expressions for} the unperturbed electron 
and phonon Green's functions. The self-energy is then analytically continued to the real frequency 
axis \citep{Mahan2000}, and only the diagonal terms are retained, as in Eq.~\eqref{eq:sigmapw}. 
The electron-phonon matrix element $g$ is defined as:
\begin{equation} \label{eq:eph-matel}
 g_{mn\nu}(\bk,\bq)=\langle\psi_{m\bk+\bq}|{\Delta}_{\bq\nu}V_\textup{KS}|\psi_{n\bk}\rangle .
\end{equation}
and it contains the variation of the self-consistent {Kohn-Sham (KS)\cite{KS1965}} potential $V_\textup{KS}$ 
with respect to a phonon perturbation. The umklapp processes are included by letting $\bk+\bq$ fall outside 
the first Brillouin zone and folding it back with a reciprocal lattice vector ${\bf G}$. The definition in 
Eq.~\eqref{eq:eph-matel} corresponds to taking the bare Coulomb potential between the electrons and the nuclei 
screened by the electronic dielectric function $\epsilon_{{\bf G},{\bf G'}} (\bq,\omega)$. In principle the 
matrix element should be frequency dependent, however in \textit{ab initio} calculations it is taken to be 
static, following the adiabatic approximation of standard density-functional theory (DFT). In 
Sec.~\ref{sec:ph-sat-calc} we will discuss how going beyond this approximation is needed when describing 
polarons in ARPES spectra. In practical calculations Eq.~\eqref{eq:eph-matel} is evaluated using 
density-functional perturbation theory (DFPT) by determining the linear variation of the self-consistent 
Kohn-Sham potential. A rigorous discussion of the calculation of the DFPT screening as compared to the 
many-body random-phase approximation (RPA) can be found for example in \citep{Marini2015}. 

\subsection{The electron spectral function} \label{sec:A}

The calculation of the self-energies defined by Eqs.~\eqref{eq:sigmapw} and \eqref{eq:sigma-eph} 
constitutes the first step towards the description of satellites from first principles. Details regarding 
the numerical evaluation of these expressions have been thoroughly reported for instance in \citep{Marini2009} 
and \citep{Ponce2016} and will not be discussed here. Once the electron self-energy {$\Sigma_{n\bk}(\omega)$} 
is known, the spectral function is obtained by combining Eq.~\eqref{eq:A} with the Dyson's equation 
$G_{n\bk} = [\hbar\omega - \ve_{n\bk} -\Sigma_{n\bk} (\omega)]^{-1}$, which yields: 
\begin{equation}\label{eq:Adiag}
A(\bk,\omega) = -\frac{1}{\pi}\sum_n \frac{ {\rm Im} \Sigma_{n\bk}(\omega) }
{ [ \hbar\omega - \ve_{n\bk} -  {\rm Re} \Sigma_{n\bk}(\omega)]^2 
+ [ {\rm Im} \Sigma_{n\bk}(\omega)]^2} 
\end{equation}
The spectral function exhibits sharp peaks whenever the first term in the denominator of Eq.~\eqref{eq:Adiag} 
$[ \hbar\omega - \ve_{n\bk} -  {\rm Re} \Sigma_{n\bk}(\omega)]$ vanishes or has a minimum. In particular, 
quasiparticle peaks in the spectral function arise at the energies $\hbar\omega= \ve_{n\bk} + 
Z_{n\bk} {\rm Re} \Sigma_{n\bk}(\ve_{n\bk})$, where 
$Z_{n\bk} = [1-\left.\partial {\rm Re} \Sigma_{n\bk}(\omega) / \partial \omega \right|_
{\omega = \ve_{n\bk}}]^{-1}$ is the quasiparticle weight. If the Bloch single-particle energies 
$\ve_{n\bk}$ are obtained from a DFT calculation, Eq.~\eqref{eq:Adiag} should be modified to avoid 
double counting of the exchange-correlation \citep{AULBUR2000}.

A more suitable framework for the evaluation of satellites in photoemission is provided by the cumulant 
expansion approach. The cumulant expansion is an alternative formulation of the (retarded) single-particle 
Green's function which is in principle exact. The Green's function is expressed in the form 
\citep{Gumhalter/2016/PRB,Kas2014}: 
\begin{equation} \label{eq:Gcum}
G_{n{\bk}}(t) = i \theta(t) {\rm exp} [-i (\ve_{n{\bk}}-i\eta ) t /\hbar  + C_{n{\bk}}(t)] ,
\end{equation}
where we introduced the cumulant function $C_{n{\bf k}}(t)$ which is defined by: 
\begin{equation}\label{eq:Cbranko}
C_{n{\bk}}(t) = \frac{1}{\hbar\pi} \int d\omega \,{\rm Im}\,\Sigma_{n{\bk}} (\ve_{n{\bk}}/\hbar - \omega )  
\frac{ 1 - e^{i\omega t} + i\omega t }{\omega^2}
\end{equation}
In practice, the {spectral function obtained from Eqs. \eqref{eq:A}, \eqref{eq:Gcum} and \eqref{eq:Cbranko} } 
can be recast into a form that is more suitable for numerical calculations 
\citep{Ferdi1996,Aryasetiawan1998,Verdi2017}:
\begin{multline}\label{eq.exponent}
A(\bk,\omega)={\sum}_n \left[ 1+
  A_{n\bk}^{\rm S1}(\omega) \ast \phantom{\frac{1}{2}} \right. \\
  \left. + \frac{1}{2} A_{n\bk}^{\rm S1}(\omega) \ast  A_{n\bk}^{\rm S1}(\omega) \ast + \cdots \right]
   A_{n\bk}^{\rm{QP}}(\omega).  
\end{multline}
Here we introduced the following quantities: 
 \begin{align} 
  A^{\rm{QP}}(\omega)&= \frac{e^{\,{\rm Re}\varSigma'(\ve/\hbar)}}{\pi}\frac{|{\rm Im}\varSigma(\ve/\hbar)|}
    {[\hbar\omega-\ve-{\rm Re}\varSigma(\ve/\hbar)]^2+[{\rm Im}\varSigma(\ve/\hbar)]^2}, \nonumber \\[5pt]
  A^{\rm S1}(\omega)
  &= -\frac{1}{\pi} \frac{ {\rm Im}\varSigma(\ve/\hbar+\omega) -{\rm Im}\varSigma(\ve/\hbar) -
  \hbar\omega\,{\rm Im}\varSigma'(\ve/\hbar)} {(\hbar\omega)^2}, \nonumber 
 \end{align}
where we omitted the dependence on $n$ and ${\bf k}$, and the prime symbol denotes the first derivative.
In the limit of a localized electron interacting with a plasmon bath, one may show that 
Eq.~(\ref{eq.exponent}) reduces to the exact solution of the localized electron model given by 
Eq.~(\ref{eq:Cmod}). The application of this formalism to core and valence excitations of crystalline 
solids, on the other hand, involves several approximations such as neglecting {\it recoil} effects, 
that is the correlations between successive boson emission and reabsorption events. A detailed discussion 
of the range of validity of the cumulant expansion has been reported, for instance, 
in \citep{Hedin1980,Gumhalter/2016/PRB,Kas2014,Sky2015}.

As discussed by \citep{Holm1997}, the cumulant expansion has the advantage of introducing additional 
crossing and non-crossing Feynman diagrams that are neglected in the standard $GW$ and Migdal
approximation for the self-energy, and it results in an improved description of the electron-plasmon 
and electron-phonon interactions. The {\it ab initio} cumulant expansion approach is based on the 
evaluation of Eq.~\eqref{eq.exponent} employing either the $GW$ or Migdal self-energy. As
discussed in Sec.~\ref{sec:pl-ph-sat}, this formalisms also lends itself to describe the 
 
The self-consistent solution of the Dyson's equation could in principle provide an alternative 
route to include additional diagrams beyond the $GW$/Migdal approximation in the Green's function.
However, while self-consistent $GW$ has been shown to systematically improve the description of the 
quasiparticle energies of molecules and solids \citep{Caruso2013,CarusoJCTC,Kutepov2012}, 
the study of satellites has revealed that self-consistency leads to an unphysical renormalization 
of the satellite intensity which, ultimately, is expected to deteriorate the agreement with 
experiment \citep{Holm1998}. Additional first-principles investigations would be needed to 
further explore this aspect. 

\subsection{Plasmon satellites} \label{sec:plas-sat}

The concept of plasmons, collective fluctuations of the electron density, can be introduced based on 
a simple model of carrier dynamics for a homogeneous system (that is, a system characterized by a 
homogeneous electron density and a positively-charged ionic background) in which the quantum-mechanical 
character of the electrons is ignored. If an external perturbation as, for example, a homogeneous electric 
field is present, a displacement ${\bf x}$ of the electron density with respect to the positively charged 
ionic background is induced. The displaced electron density then generates an induced polarization 
${\bf P}= -n e {\bf x}$, where $n$ is the average electron density and $e$ the electron charge, and 
an electric field ${\bf E} = -{\bf P}/\varepsilon_0$. 
Using Newton's law $m\ddot{\bf x}=-e{\bf E}$ with $m$ the electron mass, the classical equation of motion for the 
density displacement vector ${\bf x}$ may be rewritten as $\ddot{\bf x} + n e^2{\bf x}/(\varepsilon_0 m)=0$.
This model illustrates that the classical collective dynamics of electrons in solids can be approximately 
described by a harmonic oscillator with a characteristic frequency $\omega_{\rm P} = \sqrt{n e^2/(\varepsilon_0 m)}$, 
the plasma frequency, which is independent of the perturbation and {is determined} exclusively {by} 
the intrinsic properties of the solid.

\begin{figure}[!b]
\begin{center}
\includegraphics[width=0.4\textwidth]{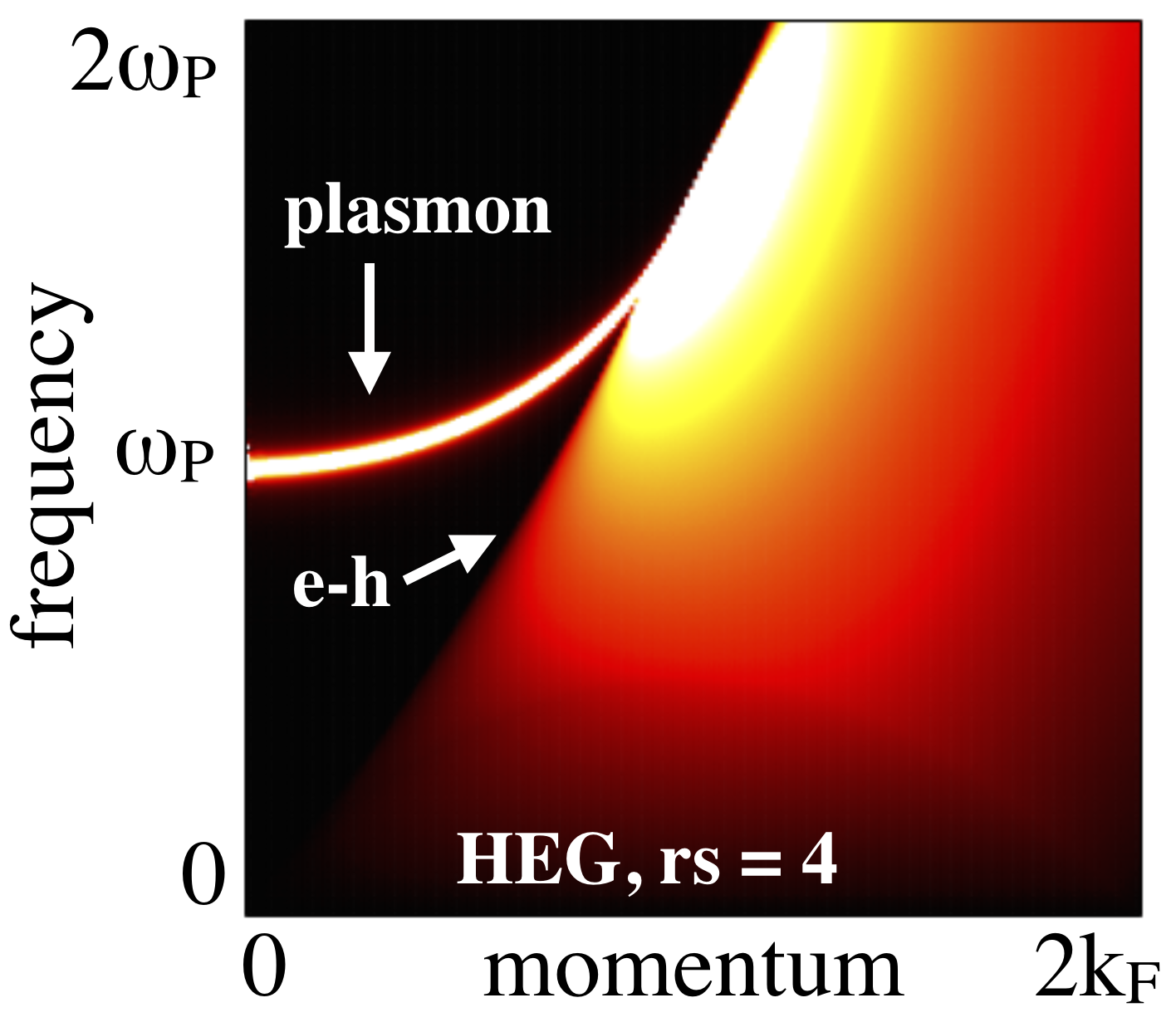}
\caption{\label{fig:lossHEG} The loss function of the HEG for a Wigner-Seitz radius $r_s=4$. 
The plasmonic structures in the loss function follow a characteristic parabolic dispersion 
which, for momenta smaller than a critical momentum $q_{\rm c}$,  is well separated from the 
continuum of electron-hole excitations (e-h). 
For $q>q_{\rm c}$, the plasmons are damped by the interaction with electron-hole pairs (Landau damping). }
\end{center}
\end{figure}

More generally, plasmons in solids may be excited at momenta $\bq$ and frequencies $\omega_{\rm P}$ which 
correspond to vanishing real part of the macroscopic dielectric function $\epsilon_{\rm M}$ and sufficiently 
small imaginary part, that is: 
\begin{equation}\label{eq:pleps}
\epsilon_{\rm M}(\bq,\omega_{\rm P}) = i\eta, 
\end{equation}
The macroscopic dielectric function  $\epsilon_{\rm M}$ is related to the microscopic dielectric function 
$\epsilon$ via $\epsilon_{\rm M}^{-1} (\bq,\omega) = [\epsilon_{{\bf G},{\bf G'}} (\bq,\omega) ]^{-1}_{{\bf G}
={\bf G'}=0}$, with ${\bf G}$ and ${\bf G'}$ reciprocal lattice vectors and $\bq$ in the first Brillouin zone. 
Whenever the condition expressed by Eq.~(\ref{eq:pleps}) is satisfied, the system may support collective 
charge fluctuations even in the absence of an external driving field. In practice, the possibility of 
exciting plasmons is reflected by the emergence of sharp peaks in the loss function 
$L(\bq,\omega) = {\rm Im}\,[\epsilon_{\rm M}^{-1} (\bq,\omega)]$ 
at the momenta and frequencies at which the macroscopic dielectric function $\epsilon_{\rm M}$ vanishes. 
The plasmon peaks in the loss function exhibit well defined energy-momentum dispersion relations. 
These structures are exemplified in Fig.~\ref{fig:lossHEG} for the loss function of the homogeneous 
electron gas {(HEG)}. If one neglects local-field effects, and thus assumes that the macroscopic 
and microscopic dielectric functions coincide, the plasmon energy is obtained by seeking the frequencies 
that satisfy the condition $v({\bf q})^{-1}= {\rm Re}\,\chi_0({\bf q},\omega)$. 
For the HEG in the long-wavelength limit (${\bq}\rightarrow0$), this condition yields again 
the result $\omega_{\rm P}=\sqrt{e^2 n/(\varepsilon_0 m)}$. 

The inspection of Eqs.~(\ref{eq:pleps}) and (\ref{eq:sigmapw}) reveals that when the condition for the 
excitation of plasmons is satisfied, the screened Coulomb interaction $W_{{\bf G},{\bf G'}}(\bq,\omega) = 
v_{\bf G}(\bq) [\epsilon_{{\bf G},{\bf G'}} (\bq,\omega)]^{-1}$ exhibits a pole at the plasmon energy. 
Correspondingly, one expects the $GW$ self-energy to encode information regarding electron-plasmon interaction.

To examine in more detail the inclusion of electron-plasmon coupling effects in the $GW$ self-energy, 
we discuss below its connection with the electron-boson coupling model introduced in Sec.~\ref{sec:model}.
Using the condition given in Eq.~(\ref{eq:pleps}) in combination with Eq.~(\ref{eq:sigmapw}), the 
plasmonic contribution to the screened Coulomb interaction $W$ can be disentangled from the other 
electronic contributions to the screening, such as electron-hole pairs. This idea, initially introduced 
for the homogeneous electron gas \citep{Lundqvist1967} and subsequently generalized to semiconductors 
\citep{Caruso/2016/PRB}, allows one to define a self-energy which stems exclusively from the coupling 
between electrons and plasmons. The resulting electron-plasmon self-energy can be recast into the following 
form \citep{Caruso/2016/PRB}:
  \begin{align}
  \Sigma^{\rm eP}_{n\bk} =& \int\!\frac{d\bq}{\Omega_{\rm BZ}}\sum_m
     |g^{\rm eP}_{mn}(\bk,\bq)|^2 \nonumber \\
  & \times \left[ \frac { n_\bq + f_{m\bk+\bq} }
  {\varepsilon_{n\bk} - \varepsilon_{m\bk+\bq} + \hbar\omega_{\rm P}(\bq) + i\eta } \right. \nonumber\\ 
  & \left. + \frac { n_\bq + 1 - f_{m\bk+\bq} }{\varepsilon_{n\bk} - \varepsilon_{m\bk+\bq} - 
  \hbar\omega_{\rm P}(\bq) + i\eta  } \right], \label{eq:sigma}
  \end{align}
where the coefficients $g^{\rm eP}_{mn}(\bk,\bq)$ are the electron-plasmon scattering matrix elements
between the initial state $\psi_{n\bk}$ and the final state $\psi_{m\bk+\bq}$ and are given by:
  \begin{equation}\label{eq:gs}
  g^{\rm eP}_{mn}({\bf k},{\bf q}) =
  \left[\frac{\varepsilon_0\Omega}{e^2 \hbar} \frac{\partial\epsilon({\bf q},\omega)}
   {\partial\omega} \right]_{\omega_{\rm P}({\bf q})}^{-\frac{1}{2}}
   \frac{1}{|{\bf q}|}\langle \psi_{m{\bf k+q}} |e^{i{\bf q}\cdot{\bf r}} | \psi_{n{\bf k}} \rangle ,
  \end{equation}
with $\Omega$ {being} the volume of the unit cell. 
Equation~\eqref{eq:sigma} has the form of an electron-boson coupling self-energy in the Migdal 
approximation [see Eq.~\eqref{eq:sigma-eph}], which may alternatively be derived from an electron-boson 
coupling Hamiltonian of the form:
\begin{equation}\label{eq:Hepl}
\hat {H}^{\rm eP} = \sum_{n m} \sum_{\bk,\bq} \, g^{\rm eP}_{nm}(\bk,\bq)
\hat c_{m\bk+\bq}^\dagger \hat c_{n\bk} (\hat b_\bq + \hat b^{\dagger}_{-\bq}).
\end{equation}
Here $\hat b^{\dagger}_{-\bq}$ ($\hat b_\bq$) and $\hat c_{m\bk+\bq}^\dagger$ ($\hat c_{n\bk}$)
are the boson and fermion creation (destruction) operators, respectively. The localized electron model 
of Eq.~\eqref{eq:Heb} is recovered from Eq.~\eqref{eq:Hepl} by (i) replacing the Bloch energies 
$\ve_{n{\bk}}$ with a single non-dispersive energy and (ii) neglecting the ${\bk}$-dependence of 
the electron-boson coupling matrix elements. This result indicates that the $GW$ self-energy accounts 
for the coupling between electrons and plasmons. However, at variance with the localized electron model 
which could be solved exactly, here the electron-plasmon interaction is treated only at first-order in 
the interaction strength, which corresponds to the Migdal approximation in the ordinary electron-boson 
coupling theory.

The inclusion of electron-plasmon coupling in the $GW$ theory is reflected by the emergence of plasmon 
satellites in the spectral function, which are analogous to the satellite features discussed in 
Sec.~\ref{sec:model}. In fact, in the presence of plasmons, the frequency dependence of the self-energy 
typically presents a pole, which may produce additional satellite structures in the spectral function 
signaling the coupling to plasmons.

Two clear shortcomings emerge when evaluating the spectral function within the $GW$ approximation, and 
limit its predictive power for the description of satellites in PES: (i) the energy difference between 
the satellite and the quasiparticle peak is typically overestimated by a factor of 1.5 with respect to 
photoemission experiments, and (ii) the $GW$ approximation may erroneously predict the formation of 
spurious {\it plasmaron} peaks, which stem from additional solutions of the quasiparticle equation and 
often result in an overestimation of satellite intensities. The concept of plasmaron was initially 
introduced by \citep{Lundqvist1967} as a new quasiparticle state emerging from the strong coupling 
between electrons and plasmons. Later studies, however, revealed that plasmaron peaks are an artifact 
of the $GW$ approximation and, in fact, they disappear when one resorts to a more accurate level of 
theory \citep{Langreth1970}. 
These issues can be illustrated by using a simplified model for the $GW$ self-energy:  
$\Sigma(\omega) = \alpha (\hbar \omega - \ve + \omega_P + i\eta )^{-1}$.
This expression is derived from Eq.~(\ref{eq:sigmapw}) by (i) assuming non-dispersive electron energies, 
(ii) replacing the oscillator strengths by $\delta$ functions, (iii) using a plasmon-pole model 
for the dielectric function in the form $\epsilon^{-1}(\omega) = 1 + \tilde\omega / (\omega^2 - 
\omega_{\rm P}^2 + i\eta)$, and (iv) carrying out the frequency integration analytically. 

\begin{figure}
\includegraphics[width=0.48\textwidth]{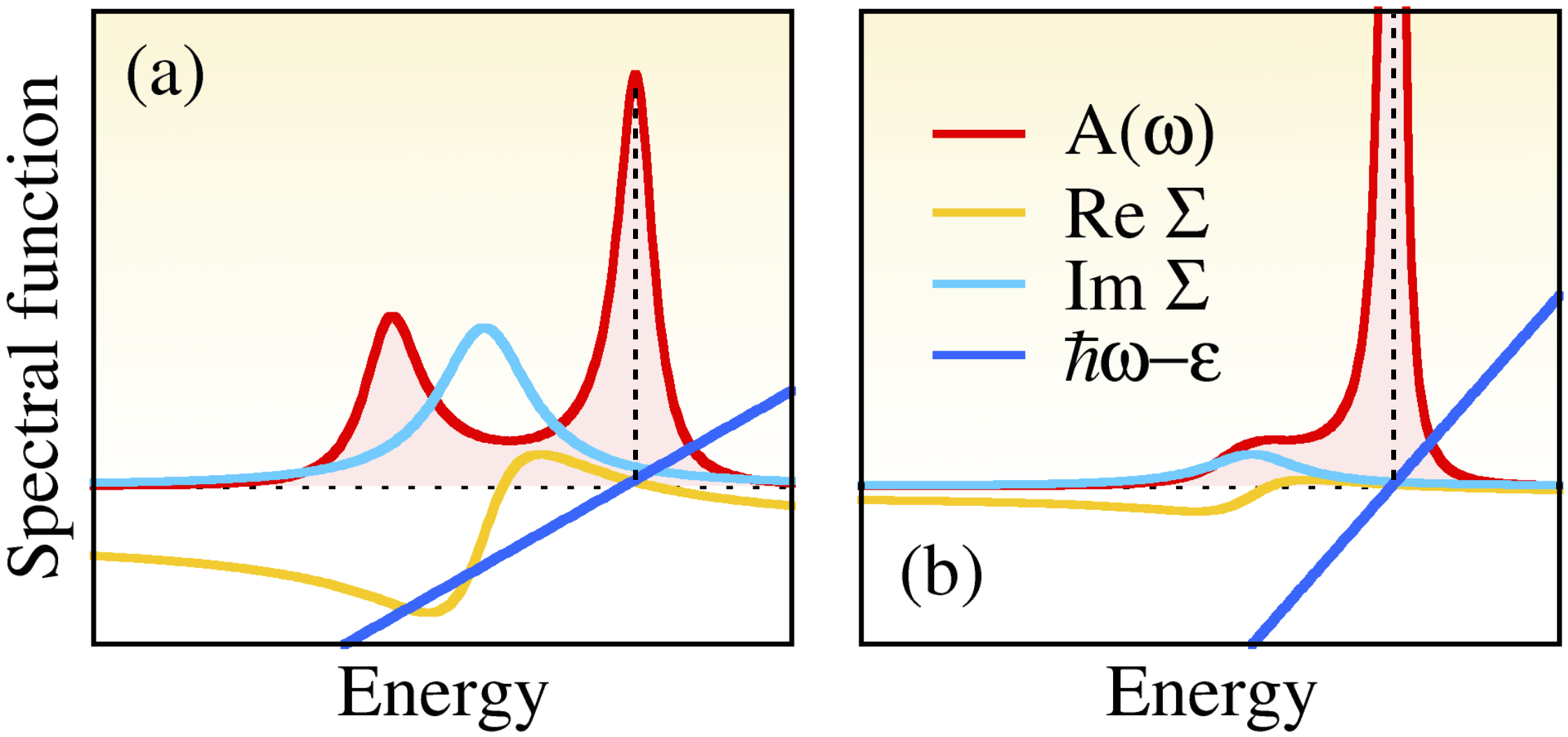}
\caption{\label{fig:sigmod}  Spectral function for a model self-energy in the strong (a) and weak (b) 
coupling regime. Quasiparticle peaks are marked by vertical dashed lines and {correspond to} the 
intersection between {${\rm Re}\Sigma(\omega)$ and $\hbar\omega-\ve$.} 
}
\end{figure}

As shown in Fig.~\ref{fig:sigmod}, the self-energy exhibits a pole at frequencies around 
$\omega = \hbar^{-1}(\ve - \hbar\omega_{\rm P})$, which may lead to additional unphysical solutions 
of the quasiparticle equation when $\hbar\omega -\e = {\it Re}\,\Sigma(\omega)$ as shown in panel (a), 
or to a minimum in $\hbar\omega -\e - {\it Re}\,\Sigma(\omega)$ resulting in a weak satellite, as shown 
in panel (b). In both cases the spectral function is characterized by the emergence of satellites, however 
their {binding} energy is blue-shifted with respect to the energy $\e-\hbar\omega_{\rm P}$ at which 
satellites are typically observed in PES experiment. 
As we will discuss in Sec.~\ref{sec:pla}, the combination of the $GW$ approximation with the cumulant 
expansion approach ($GW$+C) allows to successfully address these issues and recover an energy separation 
between satellite and quasiparticle peaks that agrees well with PES measurements for a broad class of materials. 

\subsection{Polaron satellites} \label{sec:ph-sat}

Similarly to the case of the interaction with plasmons, the coupling between electrons and phonons 
may give rise to satellite structures in the spectral function of semiconductors and insulators, which 
are the signature of the dressing of the electronic quasiparticles as \textit{polarons}. The formation 
of polarons is typically linked to the polarization of the lattice induced by longitudinal optical (LO) 
phonons. In other words, in polar semiconductors and insulators the fluctuations of the ionic positions 
corresponding to LO phonons at long wavelength generate macroscopic electric fields which can couple 
strongly to electrons and holes. This long-range interaction is known as Fr\"ohlich coupling. The 
Fr\"ohlich model strictly describes the interaction of a conduction electron in a parabolic band with 
LO phonons of constant energy $\hbar\omega_\textup{LO}$, in an isotropic and uniform medium 
\citep{Frohlich1954}. Under these assumptions, the electron-phonon matrix element does not depend 
on the band index and electron momentum, and it takes the form:
\begin{equation} \label{eq:Frohl-g}
 g_\textup{F}(\bq)=\frac{i}{|\bq|}\left[\frac{e^2}{4\pi\varepsilon_0}\frac{4\pi}{\Omega}
 \frac{\hbar\omega_\textup{LO}}{2} \left(\frac{1}{\epsilon_\infty} -\frac{1}{\epsilon_{\rm s}}\right)\right]^{1/2}\!,
\end{equation}
where $\epsilon_{\rm s}$ is the total static permittivity (or dielectric constant) and $\epsilon_\infty$ 
is the optical dielectric constant, that is $\epsilon_M(\bq=0,\omega=0)$. 
The matrix element in Eq.~\eqref{eq:Frohl-g} is often expressed in terms of a dimensionless parameter 
$\alpha$ which is referred to as the \textit{Fr\"ohlich coupling constant}: 
\begin{equation} \label{eq:Frohl-alpha}
 \alpha=\frac{e^2}{\hbar}\left(\frac{m_\textup{b}}{2\hbar\omega_\textup{LO}}\right)^{1/2}\left(\frac{1}
 {\epsilon_\infty}- \frac{1}{\epsilon_{\rm s}}\right)\!,
\end{equation}
with $m_\textup{b}$ the band effective mass of the conduction electron. The Fr\"ohlich Hamiltonian has the 
form in Eq.~\eqref{eq:Hepl} after substituting $g^{\rm eP}_{mn}$ with $g_{\rm F}\delta_{mn}$, and it is thus 
historically representative of the general problem of a fermionic particle interacting with a boson field. 
Depending on the value of $\alpha$, i.e. on the strength of the coupling, the Fr\"ohlich self-energy produces 
a spectral function that usually exhibits satellite replica of the main quasiparticle band.

First-principles calculations of electron-phonon self-energies and ARPES spectra to capture polaronic 
effects are limited by the almost prohibitive computational cost of sampling the singular behavior of 
the matrix elements for small phonon wavevectors. A procedure that enables accurate calculations of 
the electron-phonon coupling in the presence of Fr\"ohlich interaction at a reduced computational cost 
has been reported by \citep{Verdi2015,Sjakste2015}. This is achieved via the separation of the long-range, 
singular part of the electron-phonon matrix element and of the short-range part. The long-range part 
$g^\mathcal{L}$ constitutes the generalization of the Fr\"ohlich matrix element to multiple, anisotropic 
electronic bands and phonon modes, and reads: 
\begin{align} \label{eq:gL}
 g^\mathcal{L}_{mn\nu}(\bk,\bq)=&\,i\,\frac{4\pi e^2}{\Omega}\sum_\kappa\left(\frac{\hbar}
 {2M_\kappa\omega_{\bq\nu}}\right)^{1/2} \nonumber \\
 &\times\!\sum_{{\bf G},\bq+{\bf G}\neq0}\frac{(\bq+{\bf G})\cdot{\bf Z}^\ast_\kappa\cdot
  {\bf e}_{\kappa\nu}(\bq)} {(\bq+{\bf G})\cdot\bm\epsilon_\infty\cdot(\bq+{\bf G})} \nonumber \\
 &\times\langle\psi_{m\bk+\bq}|
  e^{i(\bq+{\bf G})\cdot({\bf r}-\bm\tau_\kappa^0)}|\psi_{n\bk}\rangle,
\end{align}
where ${\bf Z}^\ast_\kappa$ is the Born effective charge tensor of atom $\kappa$ in the unit cell, 
$M_\kappa$ the atomic mass and ${\bf e}_{\kappa\nu}(\bq)$ a phonon eigenvector. If combined with the 
Wannier-Fourier interpolation technique of \citep{EPW2007}, Eq.~\eqref{eq:gL} enables accurate calculations 
of polaron satellites.

As in the case of the $GW$ method, the calculation of the spectral function including electron-phonon 
coupling in the Migdal approximation suffers from two main shortcomings. First, it produces only a 
single polaronic satellite rather than a Lang-Firsov series as shown by the model of Eqs.~\eqref{eq:Heb} 
and \eqref{eq:Cmod}, and as measured in experiments. Second, its energy separation from the main 
quasiparticle peak is larger than the characteristic LO phonon energy. As {we will illustrate in 
Sec.~\ref{sec:pla} and \ref{sec:ph-sat-calc}}, the cumulant expansion method can successfully be 
employed to improve the description of {satellites. While this method has been mostly used in 
combination with the $GW$ approximation to study plasmon satellites, it} can also naturally be applied 
in the context of polaronic systems, since the theory stems from the exact solution of an electron-boson 
coupling Hamiltonian of the Fr\"ohlich type \citep{Langreth1970,Rehr2014}. The formalism corresponds 
to the one presented in Sec.~\ref{sec:A}, with {the Migdal electron-phonon self-energy used as a seed}. 

\section{Plasmon satellites in metals and semiconductors} \label{sec:pla}

\begin{figure}
\begin{center}
\includegraphics[width=.47\textwidth]{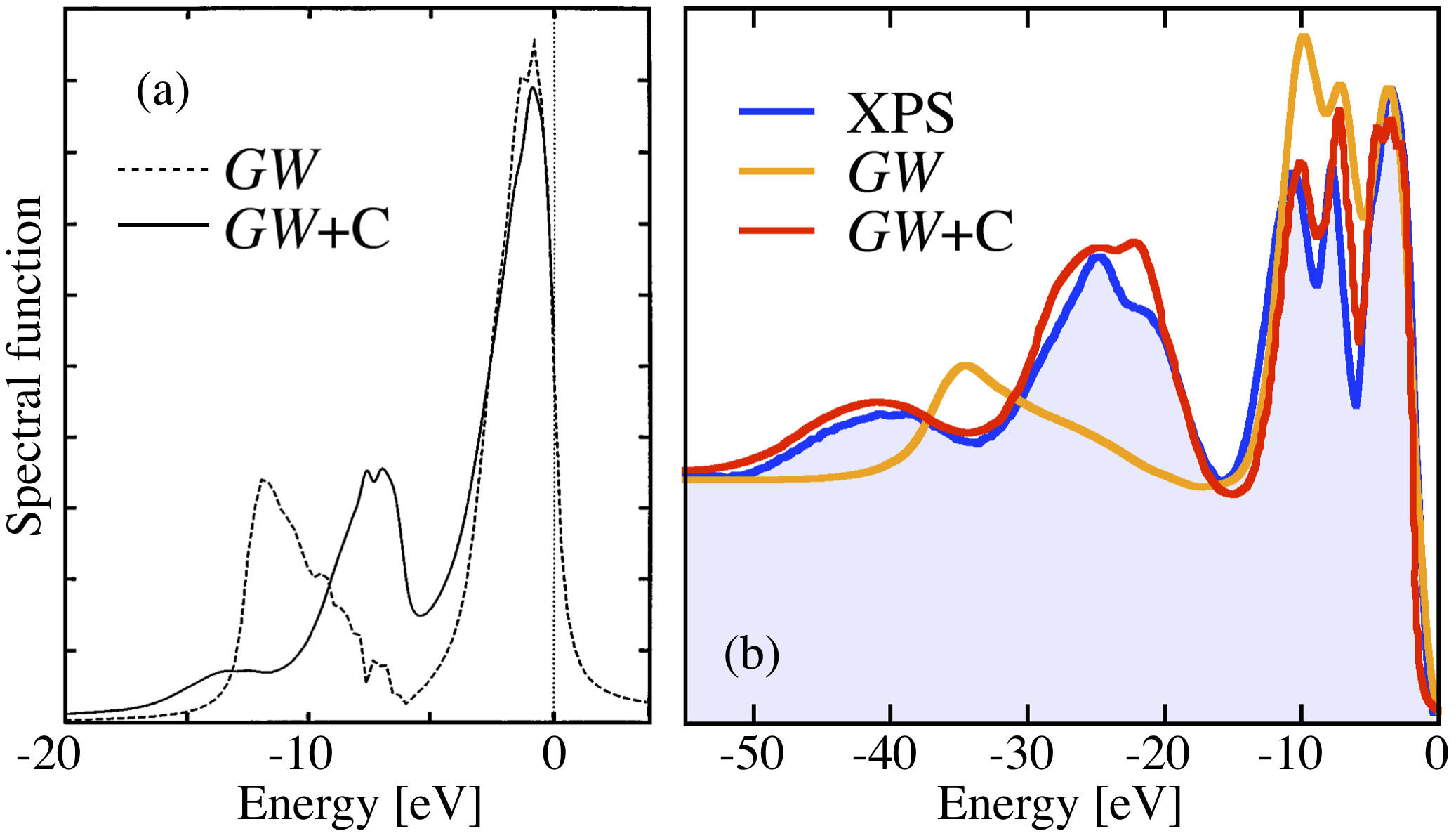}
\caption{ \label{fig:early} 
(a) Cumulant expansion for Na. 
(b) Cumulant expansion for silicon. 
Reproduced with permission from \citep{Ferdi1996,Guzzo2011}.
}
\end{center}
\end{figure}

First-principles calculations of plasmon satellites based on the $GW$+C approach have first been performed 
by \citep{Ferdi1996} for metallic sodium and aluminum. The integrated photoemission spectroscopy experiment 
on sodium by \citep{Steiner1979} revealed, besides a quasiparticle peak centered at a binding energy of 1~eV 
which corresponds to the excitation of photo-holes in the valence band, two broader and less intense 
satellite peaks blue-shifted with respect to the quasiparticle peak by 6 and 12~eV respectively. These 
energies are compatible with multiples of the plasma energy of sodium $\hbar\omega_{\rm P}\simeq5.9$~eV, 
suggesting that the satellites arise from the excitation of one and two plasmons. At variance with the 
$GW$ results, which overestimate the energy and intensity of the satellites, the spectral function of 
Na obtained from the $GW$+C approach {and shown in Fig.~\ref{fig:early}(a)} improves significantly the 
agreement with the experiment. Additionally, the $GW$+C approach lends itself to describe also processes 
in which more than one plasmon are excited, and captures the emergence of a series of satellite peaks spaced 
by the plasmon energy. On the other hand, only one satellite is obtained within the $GW$ approximation, 
reflecting the fact that multi-plasmon processes are neglected.

Subsequently, photoemission satellites have been measured in the photoemission spectra of 
graphene and the identification of these features has been supported by theoretical calculations 
of the self-energy and spectral function for linearly-dispersive bands \citep{Bostwick2010}. 
Satellites in semiconductors have first been investigated {from first principles} by \citep{Guzzo2011} 
for the case of silicon. In analogy with metals, also photoemission measurements of semiconductors may 
exhibit a series of satellites (Fig.~\ref{fig:early}(b)) with an energy separation that is compatible 
with the plasma energy. In this case, however, plasmons are generally characterized by a smaller 
oscillator strength, which is reflected by the lower intensity of the satellite peaks. 

For silicon, the $GW$ approximation yields a single satellite blue-shifted by $\sim 22$~eV 
with respect to the quasiparticle peak, which is incompatible with the plasma energy 
$\hbar\omega_{\rm P}=16.6$~eV and with the experimental observations. 
{On the other hand,} when vertex corrections are included via the $GW$+C approach the energy 
of the plasmon satellite peak is in good agreement with experiment. {Some} discrepancies between 
theory and experiment {still} remain, namely: (i) the intensity of the satellite peak is 
underestimated; (ii) the relative intensity between the different substructures of the quasiparticle 
peak differs from the experimental result; (iii) experiments present a featureless background signal 
that increases with the electron binding energy and that is not captured by theory. 

The points (i)-(iii) are directly related to the interpretation of the spectral function as a 
photoelectron current, an approximation that is typically referred to as {\it sudden approximation} 
\citep{Hufner,Damascelli2003}. In practice, the sudden approximation assumes that all electrons are 
equally likely to interact with an incoming photon, and that after photoexcitation the electrons do not 
interact further with the sample. These assumptions neglect the scattering cross-section effects due 
to the different orbital symmetries, and the additional energy losses that photo-electrons may undergo 
after emission from the initial state. The issues mentioned in (i)-(iii) could be improved by adopting 
a picture of the photoemission process that goes beyond the sudden approximation, e.g. by explicitly 
accounting for extrinsic losses, background signal, and cross-section effects. In this way, a quantitatively 
accurate description of satellites in semiconductors may be achieved \citep{Guzzo2012}.

\begin{figure*}
\includegraphics[width=0.75\textwidth]{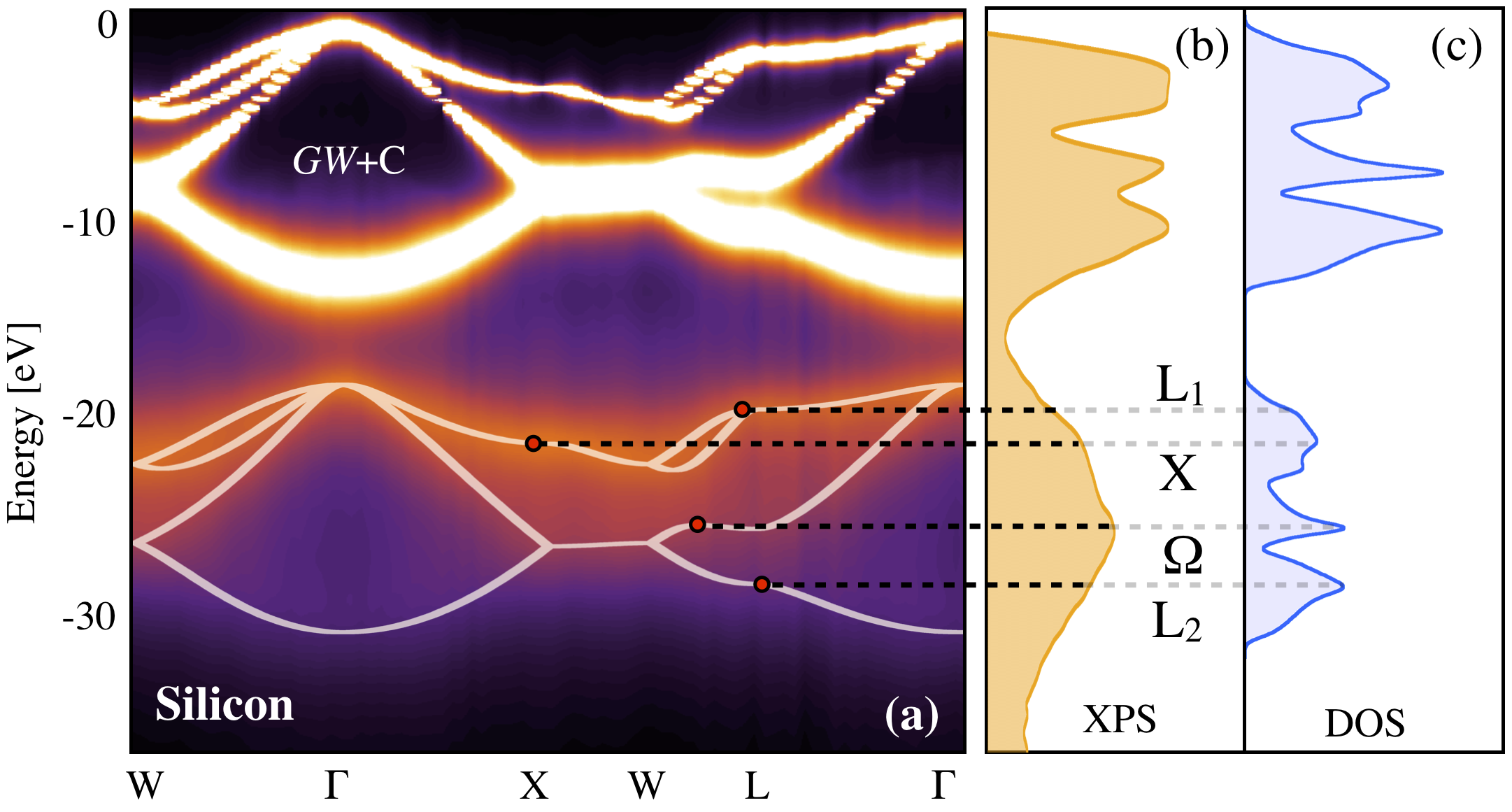}
\caption{\label{fig:plpolar} (a) Theoretical calculations of the plasmonic polaron band of silicon, 
{based on the $GW$ plus cumulant expansion approach}, adapted from \citep{Gumhalter/2016/PRB}. 
(b) Integrated X-ray photoemission spectrum {(XPS)} of silicon from \citep{Guzzo2011}. 
(c) Density of states of silicon from a density functional theory calculation alongside with 
a replica of the full DOS red-shifted by the plasmon energy $\hbar\omega_{\rm P}=16.6$~eV. 
}
\end{figure*}

Inspection of the first satellite peak in Fig.~\ref{fig:early}(b) indicates that the plasmon-induced 
spectral features of silicon are characterized by a substructure -- in this case a central peak and 
two shoulders observed in both theory and experiments -- that resembles the density of states (DOS) 
of the ordinary quasiparticle bands. To understand the origin of these features it is convenient to 
recall the concept of Van Hove singularities from the quantum theory of solids. 
The density of states $J$ for a set of Bloch electrons can be expressed as:
\begin{align}\label{eq:DOS}
J(\omega) = \frac{1}{4\pi^3}
\sum_n\int_{S(\hbar\omega)} dS_{\bf k}\frac{1}{|\nabla_{\bf k}\ve_{n{\bf k}}|} 
\end{align}
where the integral is performed over the isosurfaces in ${\bf k}$-space with energy $\hbar\omega$, denoted 
by $S(\hbar\omega)$. If for a given energy $\hbar\omega$, the isosurface $S(\hbar\omega)$ contains a crystal 
momentum for which the electron velocity vanishes ({$v_{n{\bk}}=\nabla_{\bk} \e_{n{\bk}}/\hbar=0$}), the 
divergence of the integrand in Eq.~\eqref{eq:DOS} leads to a sharp structure in $J(\omega)$, referred to 
as a Van Hove singularity. Peaks in the DOS may thus be attributed to regions of the Brillouin zone in 
which electronic bands are flat ($\nabla_{\bf k}\ve_{n{\bf k}} \simeq 0$). These structures are clearly 
visible in PES experiments of silicon for binding energies between 0 and $-15$~eV [Fig.~\ref{fig:plpolar}(b)] 
and in the DOS obtained from DFT calculations in the local density approximation [Fig.~\ref{fig:plpolar}(c)]. 
The structure of plasmon satellites measured in PES also exhibits a substructure of peaks and shoulders that 
resembles the Van Hove singularities, as it can be noted when comparing it with the DOS of the ordinary 
quasiparticle bands red-shifted by the plasmon energy [Fig.~\ref{fig:plpolar}(b)-(c)]. This suggests that 
the plasmon satellites observed in integrated PES also stem from the average over the Brillouin zone of 
spectral features that are characterized by well-defined energy-momentum dispersion relations. 

This hypothesis has been been verified by first-principles calculations of the angle-resolved spectral 
function of silicon in the $GW$+C approach \citep{Caruso/2015/PRL}, which revealed that electron-plasmon 
interaction {leads} to the emergence of {\it plasmonic polaron bands}, that is, plasmon-induced 
replica of the valence band structure of semiconductors red-shifted by the plasmon energies. These 
features are illustrated for silicon in Fig.~\ref{fig:plpolar}(a). As compared to the quasiparticle 
bands, plasmonic polaron bands are less intense due to the small oscillator strength of plasmon in 
semiconductors, and broadened out by lifetime effects. The existence of plasmonic polaron bands has 
been corroborated by further theoretical and experimental investigations of the ARPES spectrum of 
silicon for binding energies up to $40$~eV by \citep{Lischner/2015}. Overall, the dispersive character 
of plasmon-induced features in ARPES indicates that plasmon satellites in integrated PES may be 
interpreted as Van Hove singularities which arise from the flattening of the plasmonic polaron bands 
at specific regions in the Brillouin zone. 

\section{Polaron satellites in doped semiconductors} \label{sec:ph-sat-calc}

Low-energy satellites have recently been observed by ARPES experiments in doped oxides. These systems 
constitute an ideal playground for the study of polaron physics. In particular, {satellite replicas 
were} measured for $n$-doped TiO$_2$ \citep{Moser2013}{, SrTiO$_3$ \citep{Chang2010} and monolayer FeSe 
on SrTiO$_3$ \citep{Shen2014}}. Evidence of Fr\"ohlich polarons was found also from the investigation of 
two-dimensional (2D) electronic states at the surfaces or interfaces of oxides, with the most studied case 
being the 2D electron gas (2DEG) formed at the surface of SrTiO$_3$ \citep{King2014,Chen2015,Baumberger2016}. 
Other notable examples are the 2DEG at the interface between SrTiO$_3$ and LaAlO$_3$ \citep{Cancellieri2016} 
and on the surface of ZnO \citep{Yukawa2016}. The experiments also show a remarkable evolution of the 
carriers with doping concentration, from polarons to a Fermi liquid weakly coupled to phonons 
\citep{Moser2013,Baumberger2016}.

Calculations {of the spectral function} using model self-energies or the localized electron 
model of Eq.~\eqref{eq:Cmod} have been performed for some of these systems{, for example in 
\citep{Moser2013,Shen2014,King2014,Rademaker2016}}. Fully \textit{ab~initio} calculations showing 
satellite band replica were first reported for the insulating compounds MgO and LiF by \citep{Cote2015}, 
subsequently exploring also the effect of the cumulant expansion method \citep{Gonze2018}.
First-principles calculations of ARPES spectra {in doped materials} including polaronic effects 
were carried out by \citep{Verdi2017} for the prototypical case of anatase TiO$_2$ by using the methods 
presented in Sec.~\ref{sec:theo}. Given that the crystals are doped, an important element that needs to 
be taken into account when performing \textit{ab~initio} calculations is the presence of additional 
charges in the conduction band. Since the systems of interest are degenerate and present well-defined 
Fermi surfaces, doping can be treated in the rigid-band approximation, that is by placing the Fermi 
level inside the conduction or valence band of the pristine system. Moreover, the added carriers provide 
an additional source of screening of the electron-phonon interactions. This effect is critical especially 
in the case of polar coupling, where the screening of the macroscopic electric field created by the LO 
phonons can change dramatically the strength of the Fr\"ohlich interaction \citep{Mahan2000}. This aspect 
can be understood by considering the simple Thomas-Fermi screening model, which describes the static 
response of a homogeneous electron gas at small wavevectors: $\epsilon_{\rm TF}(\bq)=1+q_{\rm TF}^2/|\bq|^2$, 
with $q_{\rm TF}=\sqrt{2 e^2 n/(\varepsilon_0\epsilon_\infty E_{\rm F})}$ ($E_{\rm F}$ is the Fermi energy). 
From the wavevector dependence of $\epsilon_{\rm TF}(\bq)$ it follows immediately that the screened matrix 
element, $g_{\rm F}(\bq)/\epsilon_{\rm TF}(\bq)$, no longer exhibits a singularity at long wavelength. 
This model is valid in the adiabatic limit where the doped carriers instantaneously follow the atomic 
motion. In a more accurate description, the timescale of the electronic response is dictated by the 
plasma frequency of the doped carriers, $\omega_{\rm P}=\sqrt{ne^2/(\varepsilon_0\epsilon_\infty m_{\rm b})}$ 
in the HEG model \citep{Kittel}.

\begin{figure*} \begin{center}
\includegraphics[width=0.65\textwidth]{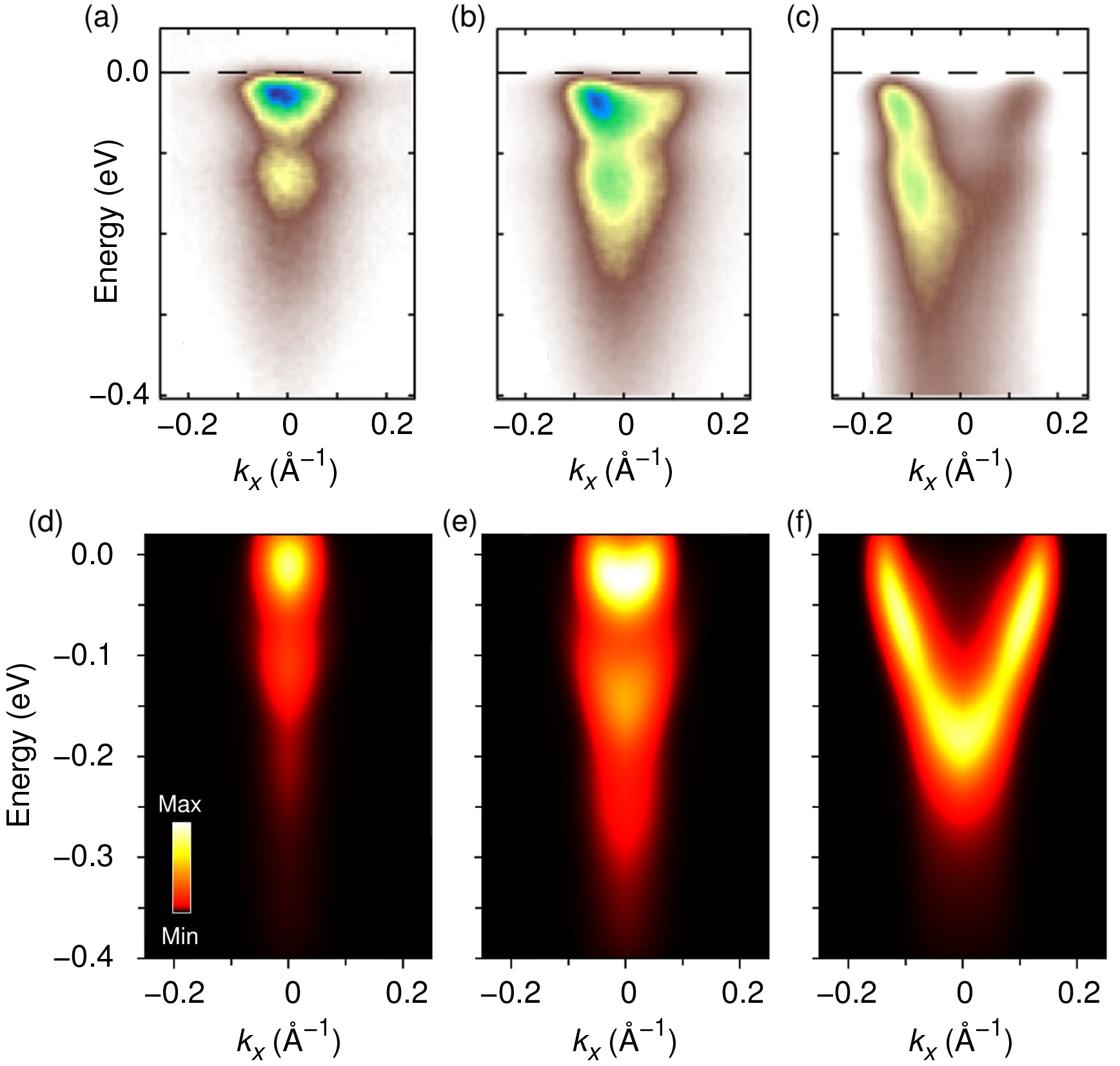}
\caption{\label{fig:spectra-tio2} ARPES spectra of $n$-doped anatase TiO$_2$ on samples with 
doping concentrations $5\times10^{18}$~cm$^{-3}$ (a), $3\times10^{19}$~cm$^{-3}$ (b) and 
$3.5\times10^{20}$~cm$^{-3}$, taken from \citep{Moser2013}. The corresponding first-principles 
spectra from \citep{Verdi2017} are shown in panels (d)-(f). 
The calculated spectral functions were multiplied by the Fermi-Dirac distribution at the experimental 
temperature ($T=20$~K) and were convoluted with Gaussian masks of widths 25~meV and 0.015~\AA$^{-1}$ 
in order to account for the experimental resolution in energy and momentum, respectively.
}
\end{center} \end{figure*}

To capture the evolution of the electron-phonon coupling and of the polaronic features with doping, 
the electron-phonon matrix element needs to be screened by the dynamical dielectric function evaluated 
at the phonon energies, that is $g_{mn\nu}^{\rm NA}(\bk,\bq)=g_{mn\nu}(\bk,\bq)/\epsilon_{\rm RPA}
(\bq,\omega_{\bq\nu}+i/\tau_{n\bk})$ \citep{Mahan2000,Verdi2017}. The superscript NA indicates that 
retardation effects are taken into account by using this non-adiabatic matrix element, and $\hbar/\tau_{n\bk}$ 
is the electron lifetime near the band edge, {which} can approximately be taken to be constant. In 
practical calculations the dynamical screening arising from the doped carriers can be computed 
analytically using the RPA dielectric function for a homogeneous electron gas with the same density $n$, 
which is known as the Lindhard function \citep{Hedin1965}.

\begin{figure} \begin{center}
\includegraphics[width=0.4\textwidth]{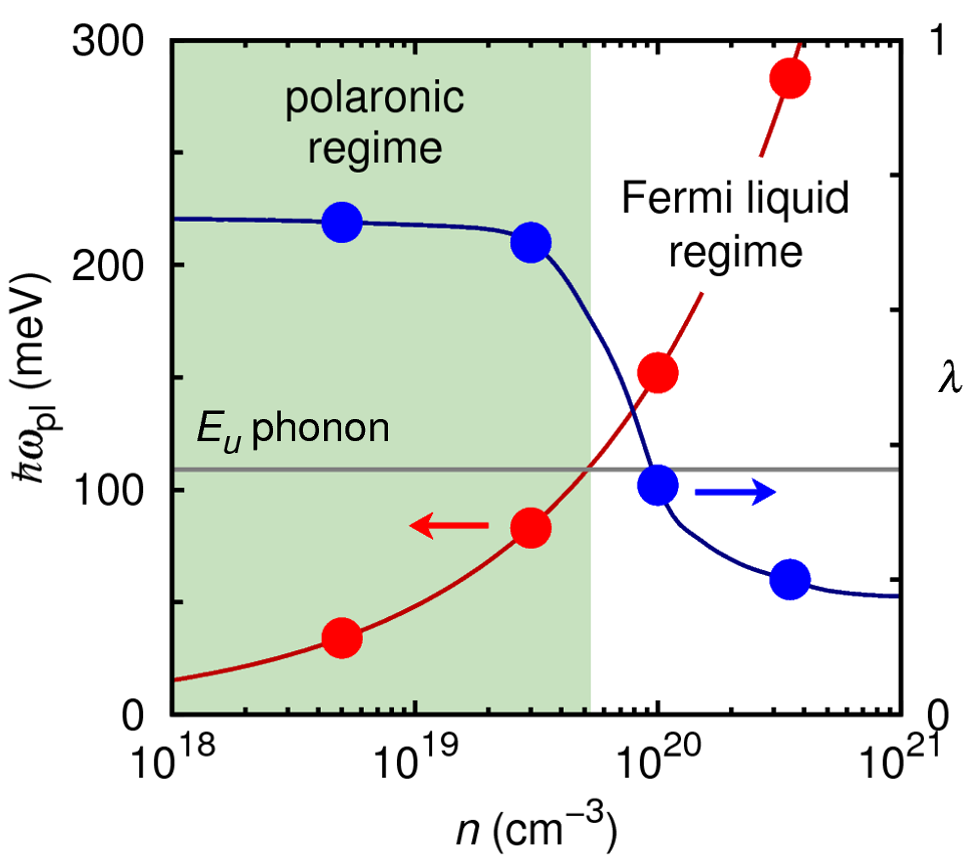}
\caption{\label{fig:crossover-tio2} Polaronic and Fermi liquid regimes in $n$-doped anatase {TiO$_2$}, from 
\citep{Verdi2017}: the red disks indicate the plasma energy at each doping level, the blue disks 
indicate the electron-phonon coupling strength $\lambda$. The blue line is a guide to the eye, while 
the red line represents the relation between the plasma energy and the doping density in the 
homogeneous electron gas. The horizontal line is the energy of the LO $E_u$ phonon of anatase {TiO$_2$}, 109 meV. 
}
\end{center} \end{figure}

In Fig.~\ref{fig:spectra-tio2} we show the ARPES spectra acquired for $n$-doped anatase TiO$_2$ 
by \citep{Moser2013}, and we compare them with the first-principles calculations performed 
by \citep{Verdi2017}. The spectra for the first two doping levels exhibit a satellite about 0.1~eV 
below the main parabolic band, and a second very dim satellite at another 0.1~eV higher binding energy. 
Since the energy separation of the band replica is compatible with the high-energy $E_u$ LO phonon of 
anatase {TiO$_2$}, these satellites were attributed to polaronic effects. At the highest doping, on the 
other hand, the satellites disappear and are replaced by band structure kinks near a binding energy of 
0.1~eV. All the spectral features and their evolution with doping are reproduced by the calculations, 
thus confirming the transition from a polaronic to a Fermi liquid picture, and demonstrating the success 
of the first-principles methods used to investigate quasiparticle spectra. From the calculated ARPES 
spectra the electron-phonon coupling strength $\lambda$ was extracted, by using the ratio between the 
Fermi velocities of the bare band and of the dressed band \citep{Mahan2000}. The results are reported 
in Fig.~\ref{fig:crossover-tio2}, together with an analysis of the energy scales at play. The study 
showed that the crossover from polarons to a weakly-coupled Fermi liquid and, correspondingly, from 
satellite replica to band structure kinks, occurs when the plasma frequency of the carriers becomes 
of the order of the LO phonon frequency. In fact, in the polaronic regime, corresponding to 
$\omega_{\rm P}<\omega_{\rm LO}$, the carriers are too slow to screen the long-range electric field 
generated by the $E_u$ phonon vibrations. In this case satellites appear in the spectra, and the 
electron-phonon coupling strength is approximately independent of doping. When $\omega_{\rm P}>
\omega_{\rm LO}$, in the Fermi liquid regime, the Fr\"ohlich coupling is strongly suppressed, with 
the polaron satellites gradually replaced by kinks. Correspondingly, the coupling strength decreases. 
This first-principles analysis indicated that the interplay between lattice vibrations and plasma 
oscillations can have a strong impact on the polaronic properties of charge carriers in doped oxides. 

\section{Hybrid plasmon-phonon satellites} \label{sec:pl-ph-sat}

\begin{figure*} \begin{center}
\includegraphics[width=0.95\textwidth]{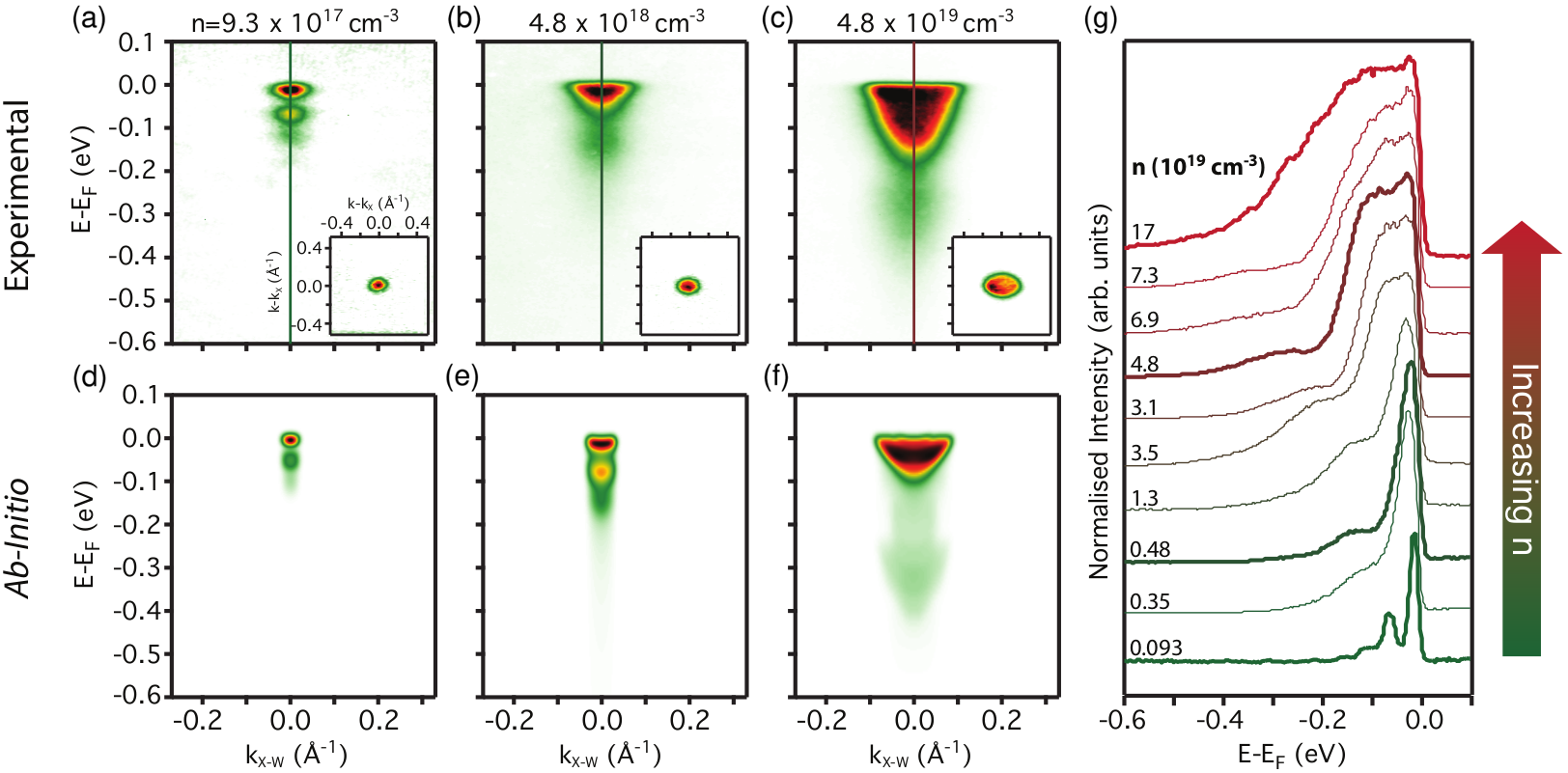}
\caption{\label{fig:euo} {(a)-(c) Measured ARPES spectra of EuO samples with increasing carrier 
concentrations as indicated on top of each panel, with the Fermi surface contours shown in the insets. 
The corresponding first-principles data are shown in (d)-(f). 
To directly compare with the experiments, the calculated spectral functions were convoluted 
with two Gaussian masks of widths 20~meV and 0.015 \AA$^{-1}$, and integrated along the out-of-plane 
direction $k_z$. (g) Measured energy distribution curves taken at $k=k_{\rm X}$ (conduction band minimum) 
for different doping levels. Figure adapted from \citep{King2018}.}
}
\end{center} \end{figure*}

Interestingly, the effects of electron-phonon and electron-plasmon interactions can be readily combined 
within first-principles calculations if the sum of the relative self-energies [Eqs.~\eqref{eq:sigma-eph} 
and \eqref{eq:sigma}] is included in the calculation of the spectral function, which can thus contain both 
plasmon and polaron satellite features. This concurrence of plasmon and polaron satellites has been observed 
experimentally in the case of the ferromagnetic semiconductor EuO, and confirmed by first principles 
calculations \citep{King2018}. Experimental ARPES spectra for Gd-doped EuO are reported in 
Fig.~\ref{fig:euo}(a)-(c) for three different doping concentrations, showing the bottom of the conduction 
band centered at the X point of the Brillouin zone. The energy distribution curves (EDCs) at the conduction 
band minimum for several dopings are reproduced in Fig.~\ref{fig:euo}(g), and they clearly show a shoulder 
peak whose energy separation with respect to the main quasiparticle band increases with carrier concentration. 

Such a satellite peak is not resolved above a carrier density $n\approx10^{20}$~cm$^{-3}$, whereas at 
low carrier concentration ($n\approx10^{18}$~cm$^{-3}$) two additional satellites can be distinguished. 
The spectra calculated with the cumulant expansion method including electron-phonon and electron-plasmon 
coupling on the same footing are presented in Fig.~\ref{fig:euo}(d)-(f), and they reproduce the features 
seen in the experiment. In particular, the calculations confirmed that for the lowest doping concentration 
the series of satellites is mainly due to phonon excitations, with the LO phonon energy of EuO being 
compatible with the peak separation energy of about 56~meV. Moving to higher dopings, the polar 
electron-phonon coupling is gradually suppressed by the free-carrier screening, and the satellite peak 
shifts to higher binding energies. Since the plasma energy increases as the square-root of the carrier density, 
this finding constitutes a fingerprint of the coupling of electrons to plasmonic excitations of the 
conduction electrons. The first-principles calculations confirmed that the renormalization of the spectral 
properties at higher dopings is due to the interplay between electron-phonon and electron-plasmon coupling, 
and that in particular the low-energy broad satellite seen in Fig.~\ref{fig:euo}(c),(f) is due to plasmonic 
excitations \citep{King2018}. 

We remark that the calculations and methods presented so far neglect the effects of mutual renormalization 
between plasmon and phonon modes, which can arise when the frequency of plasmon and phonon oscillations 
are of the same order \citep{Varga1965,Settnes2017}. The inclusion of these effects entirely from first 
principles represents one of the challenges still open in the investigation of the spectral properties 
of doped systems.

\section{Conclusions}

The emergence of satellites in photoemission spectroscopy is a {universal} manifestation of electron-boson 
interactions in solids. The origin of satellites can be ascribed to the excitation of different types of 
bosonic modes such as valence plasmons, extrinsic plasmons, or polar phonons. These spectral features have 
thus far been observed in metals, semiconductors, and highly-doped oxides. Despite the diversity of the 
physical processes that underpin the satellite formation, and the broad energy scales (from 50-100~meV up 
to 15-20~eV), many-body perturbation theory provides a unified framework for their description. In 
combination with standard approximations for the electron-electron and electron-phonon self-energies, 
the cumulant expansion approach is a powerful tool for investigating the emergence of spectral fingerprint 
of electron-boson coupling in solids.

The study of satellites in solids has thus far provided valuable insight into the many-body interactions 
between electrons, plasmons, and phonons. Recent work in this area has demonstrated that first-principles 
techniques have reached an accuracy sufficient to even precede experiments in discovering new hallmarks 
of the coupling between electrons and bosons. The emergence of satellites in photoemission spectroscopy 
is just one facet of the many effects that electron-boson interaction may induce. The recent findings 
discussed in this chapter call for a systematic investigation of the influence of low-energy plasmons 
on the formation of photoemission kinks, waterfall effects, as well as novel mechanisms of superconductive 
pairing. Furthermore, other spectroscopic techniques, such as absorption, electron energy loss, or 
time-resolved spectroscopies, provide less explored tools for investigating the coupling between electrons, 
plasmons, and phonons. Highly-doped oxides constitute a particularly exciting playground for exploring the 
influence of these phenomena on the opto-electronic properties and possible opportunities for exploiting 
these new emergent properties. In these compounds, the interplay of carriers, extrinsic plasmons, and 
polar phonons, induces complex spectral features that reflect the simultaneous excitation of plasmon and 
phonon modes and that are highly tunable via the carrier concentration.

In conclusion, the last few years have witnessed a remarkable increase in the accuracy of theoretical 
techniques for the description of the excited-state phenomena from first principles. These advances, 
alongside with a relentless increase in experimental resolution, are contributing to strengthen the 
synergy between theoretical and experimental research, providing numerous opportunities to unveil 
and understand unexplored forms of fermion-boson coupling in quantum matter.

\begin{acknowledgments}
FC thanks Matteo Guzzo for sharing the data illustrated in Fig.~\ref{fig:early}. 
The authors gratefully acknowledge support from the Leverhulme Trust (Grants PLP-2015-144 and RL-2012-001), 
the Graphene Flagship (Horizon~2020 Grant No. 785219 - GrapheneCore2), and the EPSRC (Grant No. EP/M020517/1). 
\end{acknowledgments}

\end{document}